\documentclass[manuscript, nonacm]{acmart}
\renewcommand\footnotetextcopyrightpermission[1]{}
\AtBeginDocument{}
\usepackage[T1]{fontenc}
\usepackage{amsfonts}
\usepackage{amsmath}
\usepackage{textcomp}
\usepackage{xspace}
\usepackage{pifont}
\usepackage{xspace}
\usepackage{enumitem}
\usepackage{multirow}
\usepackage{caption}
\usepackage{subcaption}
\IfFileExists{todonotes.sty}{\usepackage{todonotes}}{\providecommand{\todo}[2][]{}}
\usepackage{tikz}
\usepackage{pgfplots}
\usepackage{hyperref}
\usepackage{tikz}
\usepackage{pgfplots}
\usepackage{multicol}
\usepackage{booktabs}

\setlist{nosep}
\makeatletter
\def\input@path{{figures/}{./}}
\makeatother
\pgfplotsset{compat=1.18}
\usetikzlibrary{calc,patterns,arrows.meta,positioning,fit,backgrounds}
\usepgfplotslibrary{groupplots,statistics}

\definecolor{SuiPurple}{HTML}{7570B3}   
\definecolor{AtkRed}{HTML}{C0392B}      
\definecolor{VicGreen}{HTML}{1B9E77}    
\definecolor{AccentOrange}{HTML}{D95F02}
\definecolor{BaseGray}{HTML}{7F7F7F}    

\pgfplotsset{
  suiBar/.style={
    width=\linewidth, height=0.60\linewidth,
    ymin=0, ymax=100,
    ymajorgrids=true, grid style={draw=black!10},
    axis line style={draw=black!55}, tick style={draw=black!55},
    tick label style={font=\scriptsize}, label style={font=\small},
    title style={font=\small\bfseries},
    legend style={font=\scriptsize, draw=none, fill=none, legend cell align=left},
  },
  suiAxis/.style={
    width=\linewidth, height=0.60\linewidth,
    ymin=0, ymax=100,
    ymajorgrids=true, grid style={draw=black!10},
    axis line style={draw=black!55}, tick style={draw=black!55},
    tick label style={font=\scriptsize}, label style={font=\small},
    title style={font=\small\bfseries},
    legend style={font=\scriptsize, draw=none, fill=none},
    line width=1pt, mark size=2.2pt,
  },
}
\newcommand{\fairline}[1]{\draw[BaseGray, densely dashed, line width=1pt] (rel axis cs:0,0.5) -- (rel axis cs:1,0.5);}

\setcopyright{none}

\begin{document}
\title{Fair on the Surface: Transaction-Ordering Bias and MEV in Mysticeti DAG-based BFT Protocol}

\author{Iliya Mirzaei}
\affiliation{
  \institution{Stony Brook University}
}
\email{imirzaei@cs.stonybrook.edu}

\author{Mohammad Javad Amiri}
\affiliation{
  \institution{Stony Brook University}
}
\email{amiri@cs.stonybrook.edu}

\fancyhead{}
\begin{abstract}
Distributed systems deployed in untrustworthy environments agree on a common transaction order through Byzantine fault-tolerant (BFT) consensus protocols, and that order has real financial value in many decentralized applications: whoever influences the order can profit at other users' expense, a problem known as maximal extractable value (MEV). Mysticeti is a state-of-the-art DAG-based BFT protocol in which many validators propose blocks in parallel, and the total order is derived from the resulting DAG afterward. Mysticeti is the consensus protocol powering Sui, a production blockchain with a market capitalization of roughly \$3 billion, and it is widely believed to order transactions fairly, since many validators propose blocks in parallel and committed transactions are re-sorted by gas price before execution. We show this fairness assumption breaks down in practice, and the effect is already present on Sui's live network.
First, when vertices of the committed graph are merged into a single total order, blocks from the same round are sorted by validator index, giving lower-indexed validators a permanent head start. In our evaluation on a 13-validator network with no attacker, the lower-indexed side wins same-round ordering about 89\% of the time. Second, the gas-price re-sort intended to remove this bias uses a stable sort, so transactions paying equal fees (the common case at the reference gas price) retain the original biased order, letting an attacker profit without paying extra.
Third, a validator can amplify this advantage simply by choosing when to stay silent, a fully legitimate action that violates no protocol rule; this raises its ordering win rate above 94\%.
We measure all three exploitations, verify that Mysticeti otherwise remains resilient below the standard Byzantine fault threshold, and propose a simple fix: replace the validator-index tiebreaker with an unpredictable, per-commit random key.
\end{abstract}
\maketitle
\section{Introduction}

Distributed data management systems rely on consensus protocols to stay consistent and available despite failures~\cite{corbett2013spanner,decandia2007dynamo,bronson2013tao,kallman2008h,birman1985implementing}. Consensus protocols realize \emph{State Machine Replication} (SMR)~\cite{lamport1978time,schneider1990implementing}: every replica executes the same transactions in the same order (\emph{safety}) and every valid transaction is eventually executed (\emph{liveness}). When some replicas may behave maliciously, the protocol must be \emph{Byzantine fault-tolerant} (BFT). Classically, a single designated \emph{proposer} collects transactions, packs them into an ordered batch, and broadcasts it, so a transaction's position in that batch is its execution order. This hands the proposer a quiet power: it can choose which transactions to include and how to order them without ever violating safety or liveness.

That power becomes profit once the underlying application is holding real money~\cite{nakamoto2008bitcoin,wood2014ethereum}. In decentralized finance (DeFi), a proposer can reorder trades for gain, an exploit known as \emph{maximal extractable value} (MEV)~\cite{daian2020flash,eskandari2019sok,klages2019stability,qin2022quantifying,zhou2021high,baum2021sok,heimbach2022sok}. The textbook case is the \emph{sandwich} attack: seeing a victim's large trade on an automated market maker~\cite{xu2023sok}, the proposer places its own buy just before it and its sell just after, pocketing the price move the victim causes. On Ethereum alone, such ordering games have extracted over a billion dollars from ordinary users~\cite{mev2023chainlink}. How a consensus protocol decides transaction order is therefore a first-order fairness question, not an implementation detail.

Classical BFT protocols incur high communication costs as they rely on a single node to disseminate transactions and propose an order.
Recently, \emph{DAG-based} BFT protocols~\cite{keidar2021all,danezis2022narwhal,spiegelman2022bullshark,giridharan2024autobahn,cheng2024shardag} have been proposed to remove the bottleneck by letting every node propose blocks in parallel, forming a \emph{directed acyclic graph} (DAG) from which the total order is derived afterward. These protocols have been widely used in production systems such as Sui~\cite{sui,blackshear2024suilutris}, Aptos~\cite{aptos}, Celo~\cite{celo}, Chainlink~\cite{chainlink}, and Supra~\cite{supra}, with Sui among the largest: a market capitalization of roughly \$3 billion and more than \$150 billion in cumulative on-chain exchange volume~\cite{defillama_sui}.

Sui's consensus protocol, Mysticeti~\cite{babel2025mysticeti}, is a state-of-the-art DAG-based BFT protocol widely believed to fairly order transactions for two main reasons. First, in DAG-based BFT protocols and with all validators proposing in parallel, no single party is the sole gatekeeper of ordering. Second, and more specific to Mysticeti, after consensus commits, Sui \emph{re-sorts} the committed transactions by the fee they pay (their \emph{gas price}), so that, supposedly, whoever pays most goes first, and nobody gets a free ordering edge.

\begin{figure}[t]
  \centering
  \resizebox{\linewidth}{!}{
\begin{tikzpicture}[
  font=\small,
  box/.style={draw=black!55, rounded corners=2pt, align=center, inner sep=4pt,
              minimum height=12mm, minimum width=24mm, fill=black!3},
  sui/.style={draw=SuiPurple!70!black, fill=SuiPurple!12},
  gas/.style={draw=AccentOrange!80!black, fill=AccentOrange!12},
  outbox/.style={draw=VicGreen!60!black, fill=VicGreen!12},
  flow/.style={-{Latex[length=2mm]}, draw=black!60, line width=0.9pt},
  note/.style={font=\scriptsize, align=center, text=black!80},
]
\node[box] (in) {Pending txs\\\textcolor{AtkRed}{attacker $A$}\,,\,\textcolor{VicGreen}{victim $V$}};
\node[box, sui, right=9mm of in] (s1) {Consensus commit:\\linearize sub-DAG\\by $(\mathrm{round},\mathrm{author})$};
\node[box, gas, right=9mm of s1] (s2) {Post-consensus:\\re-sort by gas price\\(\emph{stable} sort)};
\node[box, outbox, right=9mm of s2] (o) {Execution\\order};
\draw[flow] (in) -- (s1);
\draw[flow] (s1) -- (s2);
\draw[flow] (s2) -- (o);
\node[note, below=4mm of s1, text width=32mm] (n1)
  {\textcolor{SuiPurple}{\ding{182} Low-Index Head Start}:\\$A$'s block sorts before $V$};
\node[note, below=4mm of s2, text width=32mm] (n2)
  {\textcolor{AccentOrange}{\ding{183} Tie Loophole}:\\equal fees keep $A$ first};
\node[note, above=4mm of in, text width=34mm] (n3)
  {\textcolor{AtkRed}{\ding{184} Strategic Silence}:\\$A$ speaks only on early rounds};
\draw[flow, dashed, draw=AtkRed!70] (n3) -- (in);
\draw[flow, dotted, draw=SuiPurple!80] (n1) -- (s1);
\draw[flow, dotted, draw=AccentOrange!90] (n2) -- (s2);
\end{tikzpicture}}
  \caption{How a transaction's final order is decided on Sui, and where the bias enters.
  \textbf{\textcircled{1}} Consensus produces a total order from the committed sub-DAG by (round, author);
  \textbf{\textcircled{2}} then a \emph{stable} gas-price re-sort runs on top;
  \textbf{\textcircled{3}} a validator can further tilt step~1 by choosing when to stay silent.
  All three are legitimate protocol behaviors.}
  \label{fig:hero}
\end{figure}
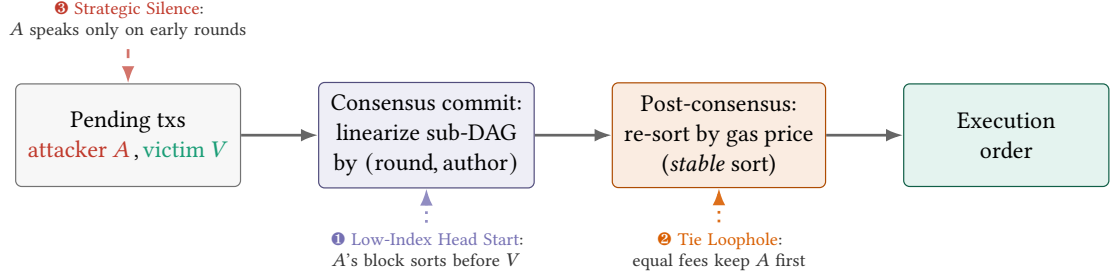

This paper shows that both reassurances are weaker than they appear, and that the resulting bias is not merely theoretical: it is currently active on Sui's production chain (mainnet), where MEV already extracts roughly \$18{,}000 a day from users~\cite{shio}, under normal operating conditions with no special configuration required. As illustrated in Figure~\ref{fig:hero}, a transaction's execution order is determined in two stages: the consensus protocol first produces a total order over the committed graph, and a gas-price re-sort is then applied on top of it. Our analysis shows that each stage independently leaks an advantage to a well-placed validator. More specifically, in a 13-validator deployment running the current Sui consensus code, we find the following opportunities for order manipulation.
\begin{itemize}
  \item \textbf{A low-index head start (\S\ref{sec:headstart}).} When the transaction blocks in the committed graph are totally ordered, same-round blocks are ordered by validator index. Hence, in the same round, the blocks generated by lower-indexed validators appear first $\approx$\,89\% of the time, with no attack running: this is the protocol's default behavior.
  \item \textbf{A tie loophole (\S\ref{sec:tieloophole}).} The gas-price re-sort is a \emph{stable} sort: on equal fees, it preserves the incoming  (biased) order. Because ordinary transactions follow the reference gas price, ties are the common case, so the head start opportunity exists for the attackers without paying any extra cost.
  \item \textbf{Strategic silence (\S\ref{sec:silence}).} A validator can amplify its head start using nothing more than a legitimate action, choosing not to broadcast on most rounds, pushing its ordering-win rate above 94\% over one-second windows.
\end{itemize}

Our investigations also show that below the Byzantine fault threshold, Mysticeti's ordering across rounds stays essentially fair. Hence, the three aforementioned order manipulation opportunities do not break consensus safety (\S\ref{sec:fix}). We further address those order manipulation opportunities with a minimal fix: replacing the validator-index tiebreak with an unpredictable per-commit key.
We implement this technique and measure it: it drops the same-round head start from 89\% to about 45\%, essentially the 50\% fair line, removing both the head start and the tie loophole gaps without touching consensus safety, liveness, or gas-price priority. Note that this fix does \emph{not} close strategic silence, whose advantage comes from round position and is bounded by causality.

The bias we study is a property of the linearization tiebreak, not of Sui: any DAG-based BFT protocol that breaks same-round ties by a fixed validator order inherits the same head start, so the fix matters for the whole family. Our finding is also distinct from prior DAG-MEV work, which studies attacker-driven front-running~\cite{zhang2024no}: the bias we measure needs no attacker, is visible only in the same-round comparison that an \emph{aggregate} all-pairs metric hides, and is live in production.

In summary, the paper makes the following contributions.
First, we identify and measure three order manipulation opportunities in the Mysticeti DAG-based BFT protocol and its production system, Sui, and verify each against the current source; two are structural flaws, and one is a legitimate-action attack.
Second, we show the advertised gas-price defense is a no-op at tied fees, which is the common case.
Finally, we give a simple, safety-preserving fix and situate it against order-fairness and MEV-mitigation work.

\section{Background}
\label{sec:prelim}

The paper focuses on the impact of MEV attacks on BFT DAG-based protocols. In this section, we give a brief background on BFT and DAG-based BFT consensus protocols. We then discuss MEV and order-fairness.

\smallskip\noindent\textbf{Byzantine fault-tolerant consensus.}
BFT protocols realize the State Machine Replication (SMR) algorithm \cite{lamport1978time,schneider1990implementing} where the system provides a replicated service whose state is mirrored across multiple deterministic replicas. At a high level, the goal of a BFT SMR protocol is to assign each client request an order in the global service history and execute it in that order \cite{singh2008bft,amiri2024bedrock}. In a BFT SMR protocol, to agree on an ordering of incoming requests, all non-faulty replicas execute the same requests in the same order ({\em safety}) and all correct requests are eventually executed ({\em liveness}). In an asynchronous system, where replicas can fail, no consensus solutions guarantee both safety and liveness (FLP result) \cite{fischer1985impossibility}.
In the Byzantine failure model, faulty nodes may exhibit arbitrary, potentially malicious behavior. In BFT consensus protocols, while there is no upper bound on the number of faulty clients, the maximum number of concurrent Byzantine replicas is assumed to be $f$ out of $3f+1$ replicas.
BFT protocols assume that a strong adversary can coordinate malicious replicas and delay communication. However, the adversary cannot subvert cryptographic assumptions.

Traditional BFT protocols, e.g., PBFT~\cite{castro1999practical} and HotStuff~\cite{yin2019hotstuff}, despite their extensive use in distributed data management systems, suffer from multiple shortcomings.
First, BFT protocols tightly couple data dissemination with the consensus routine. Specifically, they share transaction data as an integral part of the consensus routine, creating a significant performance bottleneck.
Second, most such protocols adopt a leader-based approach, where a designated leader receives client transactions, validates them, constructs a new block, and broadcasts it to all replicas. While this method ensures consistency, it also leads to resource imbalance.
Finally, the protocols suffer from a costly view-change mechanism to change the leader.

\smallskip\noindent\textbf{DAG-based BFT.} 
DAG-based consensus protocols have recently been proposed to address the shortcomings of traditional BFT protocols.
DAG-based protocols decouple transaction dissemination from the consensus routine. They further enable concurrent dissemination of transaction blocks where every validator proposes blocks continuously and in parallel. Each new block references several blocks from the previous round, recording its \emph{causal history}; the blocks and their references together form a \emph{directed acyclic graph} (DAG), a graph whose arrows always point back in time and never form a cycle. DAG-based protocols process transactions in two stages~\cite{danezis2022narwhal,keidar2021all}. In \emph{DAG construction}, validators simply gossip blocks and grow their own local copy of the graph. In \emph{ordering}, each validator independently flattens the committed part into the \emph{same} total order by applying a fixed, deterministic rule, with no extra voting messages. Different protocols differ in how they build and confirm the DAG: Narwhal/Tusk~\cite{danezis2022narwhal} and Bullshark~\cite{spiegelman2022bullshark} attach explicit certificates to blocks (a \emph{certified} DAG), whereas Mysticeti~\cite{babel2025mysticeti}, the protocol Sui runs, skips certificates and lets a block's references stand in as votes (an \emph{uncertified} DAG). They share the one feature this paper turns on: the final transaction order is produced by a \emph{linearization rule} that flattens the committed graph, and whatever tie-break that rule uses quietly decides who is ordered first (\S\ref{sec:eval}).


\smallskip\noindent\textbf{Maximal extractable value (MEV).} A blockchain's state, e.g., token balances and on-chain exchange prices, depends on the \emph{order} in which transactions execute, so whoever influences that order can extract value from it. \emph{Maximal extractable value} (MEV) is the profit a party can obtain purely by choosing the order, inclusion, or exclusion of transactions~\cite{daian2020flash,qin2022quantifying}. First studied on leader-based chains such as Ethereum~\cite{daian2020flash,eskandari2019sok}. MEV is now understood as a systemic cost paid by ordinary users~\cite{heimbach2022sok}. Its textbook forms all involve a victim trade on a decentralized exchange~\cite{xu2023sok}: \emph{front-running} places an attacker transaction just \emph{before} the victim's, to act at the old price; \emph{back-running} places one just \emph{after}, e.g., to arbitrage the price the victim moved; and a \emph{sandwich} does both around the victim, so the victim is forced to trade at the worst price~\cite{eskandari2019sok,zhou2021high}. MEV is not a flaw in any one application; it is the direct consequence of \emph{someone} getting to decide transaction order, which makes it a consensus-level fairness problem.

Daian et al.~\cite{daian2020flash} showed that control over transaction order is monetizable and that the resulting competition destabilizes consensus. Eskandari et al.~\cite{eskandari2019sok} give a taxonomy (displacement, insertion, suppression) into which our head start falls as low-cost displacement. Most MEV work targets chains with a public mempool and fee auctions; Sui has neither. Hence, the advantage instead comes from the consensus tiebreak we isolate. Closest to us, Zhang and Kate~\cite{zhang2024no} show front-running is feasible on DAG-based blockchains. We differ by pinning the advantage to a specific, deployed linearization rule, showing the advertised gas-price defense is a no-op at tied fees, and adding a legitimate-action amplifier.

\smallskip\noindent\textbf{Order-fairness.}
To mitigate order manipulation, early studies focus on censorship resistance \cite{miller2016honey}, where the goal is to ensure that correct transactions are eventually ordered, i.e., not censored. However, reordering transactions, e.g., sandwich attacks, is still possible. Similarly, reputation-based systems \cite{asayag2018fair,kokoris2018omniledger,lev2020fairledger,crain2021red} only detect unfair censorship.
Order manipulation can also be addressed by hiding transaction contents until the order is fixed (encrypted or commit-reveal mempools) or by relying on trusted hardware~\cite{stathakopoulou2021adding}. Such techniques either add latency or assume trust.
Recently, the notion of {\em order-fairness} is presented to
address the manipulation of transaction ordering \cite{zhang2020byzantine,kursawe2020wendy, kursawe2021wendy,kelkar2020order,kelkar2023themis,cachin2022quick,nagda2024rashnu}.
Order fairness ensures that the committed order of transactions respects the order in which validators actually received the transactions,
denying anyone a purely positional advantage. A foundational result is that \emph{perfect} order-fairness is impossible: the receive order seen by different honest validators can disagree in a cycle (one saw $t_1$ before $t_2$, another $t_2$ before $t_3$, a third $t_3$ before $t_1$), a Condorcet paradox that no total order can satisfy~\cite{kelkar2020order}. Protocols such as Aequitas~\cite{kelkar2020order}, Themis~\cite{kelkar2023themis}, and Rashnu~\cite{nagda2024rashnu} therefore settle for a weaker, batched notion of fairness~\cite{zhang2020byzantine,kursawe2020wendy,cachin2022quick}.
Recently, order fairness has been applied on top of DAG-based protocols, e.g., FairDAG~\cite{kang2025fairdag}, DoD~\cite{nagda2026dag}, where they layer fair ordering on multi-proposer causal designs. Finally, Mah\'e and Tucci-Piergiovanni~\cite{mahe2025order} evaluate order-fairness of DAG ledgers and conclude that the linearization rule is what determines fairness, directly supporting our thesis.
Our paper is complementary and deliberately lighter: rather than adding a fairness protocol, we show that a
\emph{deployed} DAG chain's existing linearization rule already hands out a positional advantage, and that a minimal change to that one rule removes most of it.
\section{Ordering Transactions in Mysticeti}
\label{sec:bg}

Mysticeti stores state as \emph{objects}. An \emph{owned} object has a single owner (for example, a coin in your wallet); a \emph{shared} object can be touched by anyone (for example, a trading pool). Transactions that touch only owned objects need no global order and take a fast, consensus-free path~\cite{blackshear2024suilutris}. Only transactions that touch the \emph{same shared object} (e.g., two swaps against the same pool) can conflict, so only these are sent through consensus to be put in a definite order. Ordering bias, and therefore MEV, lives entirely on this shared-object path.

Mysticeti runs on a DAG~\cite{babel2025mysticeti,keidar2021all}. Time proceeds in numbered \emph{rounds}. In each round, every validator may propose one \emph{block} (a batch of transactions) that references blocks from the previous round; those references are themselves votes for the blocks they point to, so no separate voting messages are needed. Certain blocks are designated \emph{leaders}; once a leader has enough support, it is \emph{committed}, which also commits everything the leader's block transitively references, i.e., its \emph{sub-DAG} (its causal history). To execute the result, the sub-DAG (a partial order, with many blocks left unordered relative to each other) must be turned into one total order. Mysticeti does this by \emph{linearizing}: it sorts the committed blocks by $(\mathrm{round},\mathrm{author})$ (round number first, then validator index), using a stable sort. This sort is the first place bias can enter (\S\ref{sec:headstart}).

\begin{figure}[t]
  \centering
  \resizebox{\columnwidth}{!}{
\begin{tikzpicture}[
  font=\scriptsize,
  blk/.style={draw=black!55, rounded corners=1.2pt, align=center, inner sep=1.8pt,
              minimum width=12mm, minimum height=6mm, fill=black!3},
  atk/.style={draw=AtkRed!80!black, line width=0.9pt, fill=AtkRed!8},
  vic/.style={draw=VicGreen!70!black, line width=0.9pt, fill=VicGreen!10},
  lead/.style={draw=SuiPurple!80!black, line width=0.9pt, fill=SuiPurple!12},
  ref/.style={-{Latex[length=1.3mm]}, draw=black!40, line width=0.6pt},
  flow/.style={-{Latex[length=2mm]}, draw=black!70, line width=1pt},
]
\node[blk, atk] (a0) at (0,0)      {$A_0$ ($V_0$)\\$t_1\,t_2$};
\node[blk]      (a1) at (0,-0.85)  {$A_1$ ($V_1$)\\$t_3\,t_4$};
\node[blk]      (a2) at (0,-1.70)  {$A_2$ ($V_2$)\\$t_5\,t_6$};
\node[blk, vic] (a3) at (0,-2.55)  {$A_3$ ($V_3$)\\$t_7\,t_8$};
\node[text=black!65] at (0,0.66) {round $r$};
\node[blk, lead] (L) at (2.4,-1.27) {$L$: leader\\$r{+}1$, $u_1 u_2$};
\node[text=black!65] at (2.4,0.66) {round $r{+}1$};
\foreach \a in {a0,a1,a2,a3} \draw[ref] (L) -- (\a);
\draw[flow] (3.6,-1.27) -- (5.0,-1.27);
\node[align=center, text=black!80, font=\tiny] at (4.3,-0.5) {commit $L$'s sub-DAG;\\sort by $(\mathrm{round},\mathrm{author})$};
\draw[flow, draw=black!35] (5.85,0.35) -- (5.85,-3.25);
\node[rotate=90, text=black!55, font=\tiny, anchor=south] at (5.62,-1.45) {execution order};
\node[blk, atk] (o0) at (6.65,0)      {$t_1\,t_2$};
\node[blk]      (o1) at (6.65,-0.72)  {$t_3\,t_4$};
\node[blk]      (o2) at (6.65,-1.44)  {$t_5\,t_6$};
\node[blk, vic] (o3) at (6.65,-2.16)  {$t_7\,t_8$};
\node[blk, lead](oL) at (6.65,-2.88)  {$u_1\,u_2$};
\node[text=AtkRed!85!black, right, font=\tiny] at (7.3,0) {first ($V_0$)};
\node[text=VicGreen!55!black, right, font=\tiny] at (7.3,-2.16) {$V_3$};
\node[text=SuiPurple!80!black, right, align=left, font=\tiny] at (7.3,-2.88) {$L$'s txns\\($r{+}1$)};
\end{tikzpicture}}
  \caption{Two rounds of a four-validator DAG. In round $r$ each validator proposes one block ($A_0$--$A_3$); in round $r{+}1$ the leader $L$ references
  them (references double as votes). Committing $L$ commits its causal history, the sub-DAG $\{A_0,\dots,A_3,L\}$, which is linearized by  $(\mathrm{round},\mathrm{author})$. The round number orders \emph{across} rounds ($r$ before $r{+}1$); \emph{within} round $r$, the validator index breaks the tie, so $A_0$'s transactions execute first and $A_3$'s last. If $t_1$ and $t_7$ swap the same pool, $V_0$ front-runs $V_3$ by index alone.}
  \label{fig:bgexample}
\end{figure}
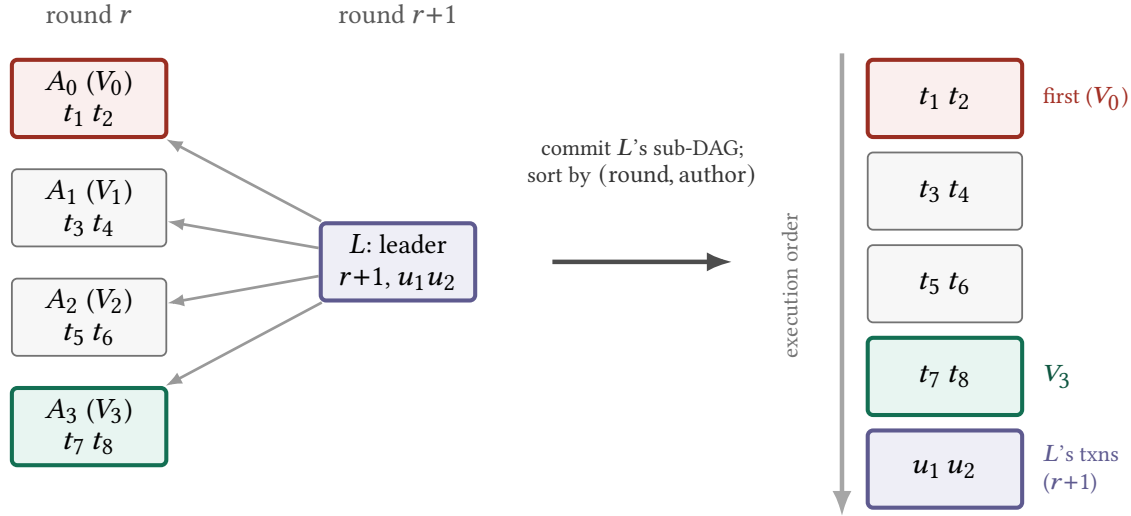

Figure~\ref{fig:bgexample} traces two rounds. In round $r$, validators $V_0$--$V_3$ each propose a block
$A_0$--$A_3$, carrying transactions. In round $r{+}1$, the leader $L$ (could be any of the validators) references those blocks (a reference doubles as a vote), and once $L$ collects enough support, it is committed, which commits its \emph{causal history}: the sub-DAG $\{A_0, A_1, A_2, A_3, L\}$. Mysticeti totally orders the blocks of this sub-DAG by $(\mathrm{round},\mathrm{author})$. The round number orders \emph{across} rounds, so all four blocks of round $r$ precede block $L$ (of round $r+1$). \emph{Within} round $r$, where the four blocks tie, the validator index alone decides the order, so $A_0$ runs first and $A_3$ last. If $t_1$ (in $A_0$) and $t_7$ (in $A_3$) swap against the same pool, $t_1$ executes first purely because $0<3$, letting $V_0$ front-run $V_3$ with no fee paid and no attack running (\S\ref{sec:headstart}).

After consensus fixes the block order, Mysticeti makes one more pass over the transactions inside the committed batch and re-sorts them by \emph{gas price} (the fee per unit of computation), highest first, before execution. Mysticeti presents this as a defense against ordering bias: pay more, go first. But every validator quotes a \emph{reference gas price} each epoch (a low, standard fee the network agrees to honor), and most ordinary transactions simply pay that reference rate. As a result, equal gas prices (ties) are the common case. What happens on a tie turns out to decide whether the re-sort is a real defense (\S\ref{sec:tieloophole}).

\section{Ordering Exploitation}
\label{sec:model}

\subsection{Threat Model}
We assume a strong but protocol-compliant adversary. A {\em rational adversary} may take any action an honest validator could legally take: which parent blocks to reference, which of several valid blocks to propose, how to fill a block, and \emph{when} to broadcast, all within the protocol's rules~\cite{zhang2024no}. It may not forge signatures, violate a validity or quorum check, rewrite committed history, or make an honest node accept anything it would normally reject. Staying inside this \emph{legitimate action set} is what makes the effects we report robust: they are not defended by any patch that hardens message handling, because no message is malformed. We assume up to $f$ rational adversary validators out of $n=3f+1$, the standard Byzantine fault tolerance (BFT) bound~\cite{pease1980reaching,dwork1988consensus}.

\subsection{The Low-Index Head Start}
\label{sec:headstart}

Recall that Mysticeti turns a committed sub-DAG into a total order by sorting blocks by (round, author). The function that does this,
\texttt{sort\_sub\_dag\_blocks} in \texttt{commit.rs}, carries the comment ``\emph{any deterministic \& stable algorithm works}.'' Any such
rule is indeed \emph{safe}. However, choosing the \emph{validator index} as the tiebreaker is not neutral: whenever two validators have a block in the same round, the one with the smaller index is always placed first. Nothing an honest higher-indexed validator does can change this.

\subsection{The Tie Loophole}
\label{sec:tieloophole}

Mysticeti addresses this ordering bias using the post-consensus gas-price re-sort: sort the committed transactions by gas price, highest first, so paying more buys priority and equal-priority transactions are, supposedly, no longer decided by consensus position. The implementation, \texttt{order\_\allowbreak by\_\allowbreak gas\_\allowbreak price} in \texttt{post\_\allowbreak consensus\_\allowbreak tx\_\allowbreak reorder.rs}, is a single call to Rust's \texttt{sort\_\allowbreak by\_\allowbreak key}. That call is a \emph{stable} sort: when two keys are equal, it preserves their input order. Here, the input order is exactly the consensus order, the head start of \S\ref{sec:headstart}. So on a tie, the re-sort changes nothing, and the low-index transaction stays first. The defense only works when one transaction pays \emph{strictly} more.

\subsection{Strategic Silence}
\label{sec:silence}

The first two findings are structural: they hold with no one misbehaving. The third is an \emph{attack} that breaks no rule.

A validator's own blocks help its overall ordering-win rate only when they land \emph{early}; a block it produces on a late round lands late and drags its average position down. The round leader is chosen in a round-robin manner, so a low-index validator leads on the \emph{first} rounds of each cycle. If such a validator simply stays silent on every round except the ones it leads, it emits only early blocks and deletes its own late ones from the comparison. Staying silent is always allowed (an honest validator with a slow link does it all the time), so no honest node rejects anything and no protocol rule is violated. In summary, this flaw encourages validators to \emph{speak only on their leader rounds}.

\section{Experimental Evaluation}
\label{sec:eval}

We now evaluate the three flaws against the current Sui source code.
We deploy the current Sui consensus code (\texttt{code/sui}) on a cluster of $n{=}13$ validators. Unless stated otherwise, we label validators $0$--$3$ as the (rational) attackers and $10$--$12$ as the victims, collect $60$ commits per run, and report the median over 5 independent runs. A run with the disabled attack represents the \emph{baseline}.

Since our metric is block-level \emph{ordering} rather than execution, transaction contents do not affect what we measure; what matters is that the input load favors no validator. Each run therefore uses a synthetic, content-neutral workload of $5n$ distinct fixed-size ($32$-byte) transactions, submitted \emph{round-robin}, so that every validator receives an equal share (five each) and the consensus layer, which never inspects a payload, treats them identically. For the two structural flaws below, which run with no attack, any ordering advantage therefore reflects the linearization rule rather than a skewed input; strategic silence keeps this same symmetric input but has the attacker choose \emph{when} to propose.

Our metric is the \emph{Attack Success Rate} (ASR)~\cite{zhang2024no}: among pairs of an attacker block $B_a$ and a victim block $B_v$, the fraction in which the attacker's block appears earlier in the committed order. We report it two ways.
First, \emph{Same-round ASR} restricts to pairs whose two blocks share a round $r$: $\Pr[B_a \text{ before } B_v \mid r(B_a) = r(B_v)]$. This isolates the linearization tiebreak.
Second, \emph{All-pairs ASR} takes every attacker--victim pair regardless of round: $\Pr[B_a \text{ before } B_v]$. A perfectly fair protocol gives $50\%$ on both, because with no systematic advantage each attacker--victim pair is equally likely to fall either way, so wins and losses cancel to a coin flip. The two metrics differ by design: across rounds, both sides produce a block every round, so those pairs cancel to $50\%$, and one can show all-pairs ASR $=\tfrac{1}{2}+\tfrac{1}{2R}$ for $R$ rounds, tending to $50\%$. The same-round metric is therefore where a per-round tiebreak shows up, and the all-pairs metric is where a genuine across-round advantage would show up.

\subsection{The Low-Index Head Start}
\label{sec:headstarteval}

\begin{figure}[t]
  \centering
  \resizebox{\columnwidth}{!}{
\begin{tikzpicture}
\begin{axis}[
  suiBar,
  ybar, bar width=13pt,
  ylabel={Same-round ASR (\%)},
  symbolic x coords={a,b,c,d,e,f},
  xtick=data,
  xticklabels={{0--3\\vs 10--12},{10--12\\vs 0--3},{5--8\\vs 0--3},{5--8\\vs 10--12},{4--5\\vs 6--7},{6--7\\vs 4--5}},
  x tick label style={font=\tiny, align=center},
  enlarge x limits=0.12,
]
\addplot+[draw=SuiPurple!70!black, fill=SuiPurple!45,
  postaction={pattern=north east lines, pattern color=black!30},
  nodes near coords, every node near coord/.append style={font=\tiny}]
  coordinates {(a,89.5)(b,10.5)(c,11.7)(d,91.7)(e,92.0)(f,8.0)};
\fairline{(a,50)(b,50)(c,50)(d,50)(e,50)(f,50)}
\node[font=\tiny, text=BaseGray, anchor=south west] at (axis cs:a,50) {50\% = fair};
\end{axis}
\end{tikzpicture}}
  \caption{The head start is caused by the validator index. We use the same data without performing any attack; we only change which validators we \emph{label} as attacker vs.\ victim. Whenever the labeled attacker has the lower index (bars 1, 4, 5) it wins $\approx$\,90\% of same-round pairs; swapping the labels flips the result to $\approx$\,10\%. The advantage tracks index position, not behavior.}
  \label{fig:headstart}
\end{figure}
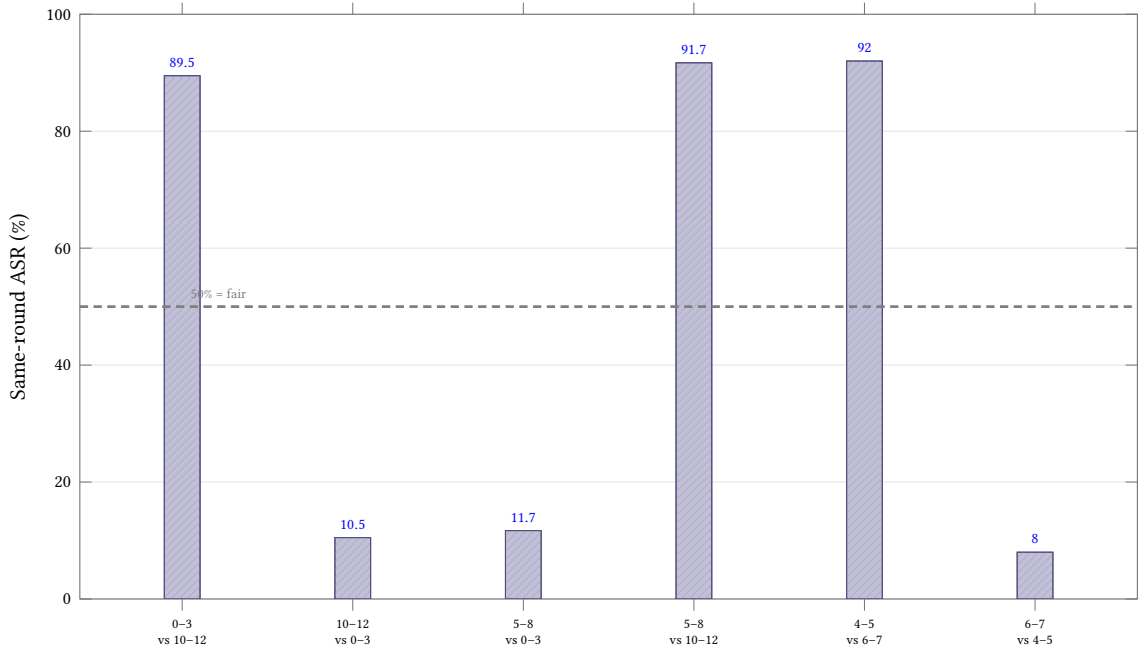

The Low-Index Head Start is a structural advantage that the Mysticeti protocol gives to the low-index validators.
To prove the advantage comes directly from index position, we run one baseline (no attack) and compute the same-round ASR under different \emph{labelings} of the same data. As shown in Figure~\ref{fig:headstart}, if we call the low-index validators the ``attacker,'' they win $\approx$\,90\% of same-round pairs; if we swap the labels and call the \emph{high}-index validators the ``attacker,'' the number flips to $\approx$\,10\%. The result mirrors around 50\%: the winner is always whoever has the lower index, even for validators two positions apart. A per-pair sweep over all $78$ validator pairs tells the same story: the lower index wins in every pair (mean 87.4\%).

The head start does not depend on a large attacker set. With only two validators per side, and a narrow index gap at that ($4$--$5$ labeled the attacker vs. $6$--$7$ the victim; the rightmost bars of Figure~\ref{fig:headstart}), the low-index side still wins $92.0\%$ of same-round pairs, dropping to $8.0\%$ once the labels are swapped. The $78$-pair sweep above is precisely the one-attacker/one-victim limit, and there too the lower-indexed validator is favored in every pair. A lone low-index validator competing against a single peer therefore already inherits the advantage; group size changes only \emph{who} shares the head start, not whether it exists.

We think this bias has not been explored until now because when we average over all pairs, the effect disappears: all-pairs ASR stays at $\approx$\,50\% because cross-round pairs cancel (\S\ref{sec:model}), and it does not grow with the number of low-index validators; it is a per-\emph{pair} property of the sort, not a coalition effect. An evaluation that reports only all-pairs ASR would therefore conclude that Mysticeti is fair. The bias lives entirely in the same-round comparison, which is basically the comparison that matters for two transactions competing to touch the same pool in the same round.

\subsection{The Tie Loophole}
\label{sec:tieloopholeeval}

\begin{figure}[t]
  \centering
  \resizebox{\columnwidth}{!}{
\begin{tikzpicture}
\begin{axis}[
  suiBar,
  ybar, bar width=22pt,
  ylabel={Attacker tx executes first (\%)},
  symbolic x coords={noreorder,tie,outbid},
  xtick=data,
  xticklabels={{No reorder\\(consensus order)},{Gas reorder,\\tied fee (common)},{Gas reorder,\\victim out-bids}},
  x tick label style={font=\tiny, align=center},
  enlarge x limits=0.30,
]
\addplot+[draw=AtkRed!70!black, fill=AtkRed!45,
  postaction={pattern=north east lines, pattern color=black!30},
  nodes near coords, every node near coord/.append style={font=\scriptsize}]
  coordinates {(noreorder,89.3)(tie,89.3)(outbid,10.7)};
\fairline{(noreorder,50)(tie,50)(outbid,50)}
\end{axis}
\end{tikzpicture}}
  \caption{The gas-price re-sort does not have any impact on ties. Share of the time the attacker's transaction executes first, for a low- vs.\ high-index pair. The re-sort protects the victim only in the rightmost case, where the victim strictly out-bids; at the common tied fee (middle), the head start passes straight through to execution.}
  \label{fig:tieloophole}
\end{figure}
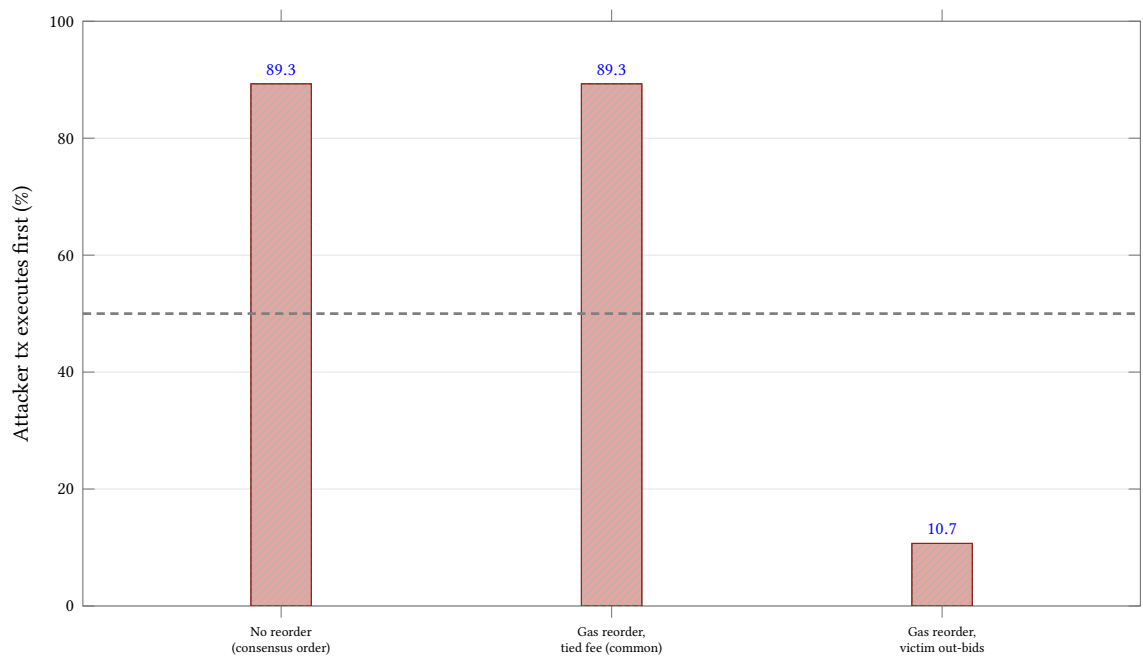

Most transactions pay the reference gas price (\S\ref{sec:bg}); as a result, ties are the common case between competing transactions.
Figure~\ref{fig:tieloophole} shows the consequence for a low- vs.\ high-index transaction pair. With no re-sort, the attacker's transaction will be executed first $\approx$\,89\% of the time (the head start). Even after applying the re-sort technique at a \emph{tied} fee, the number remains unchanged; only when the victim outbids by even one unit does the defense work. Crucially, the attacker pays no premium: the bias is exploitable at the minimum fee.

This exploitation has not been detected in the Sui blockchain, as its test for the re-sort checks only the textual form of the output transactions, which is identical for equal-fee transactions. The test therefore passes regardless of which tied transaction comes first: it checks only transaction identity, so it confirms the sort is stable without checking whether that stability is fair. The flaw is not a coding mistake in the sort; it is the assumption that a stable sort randomizes ties, when it does the opposite.

\subsection{Strategic Silence}
\label{sec:silenceeval}

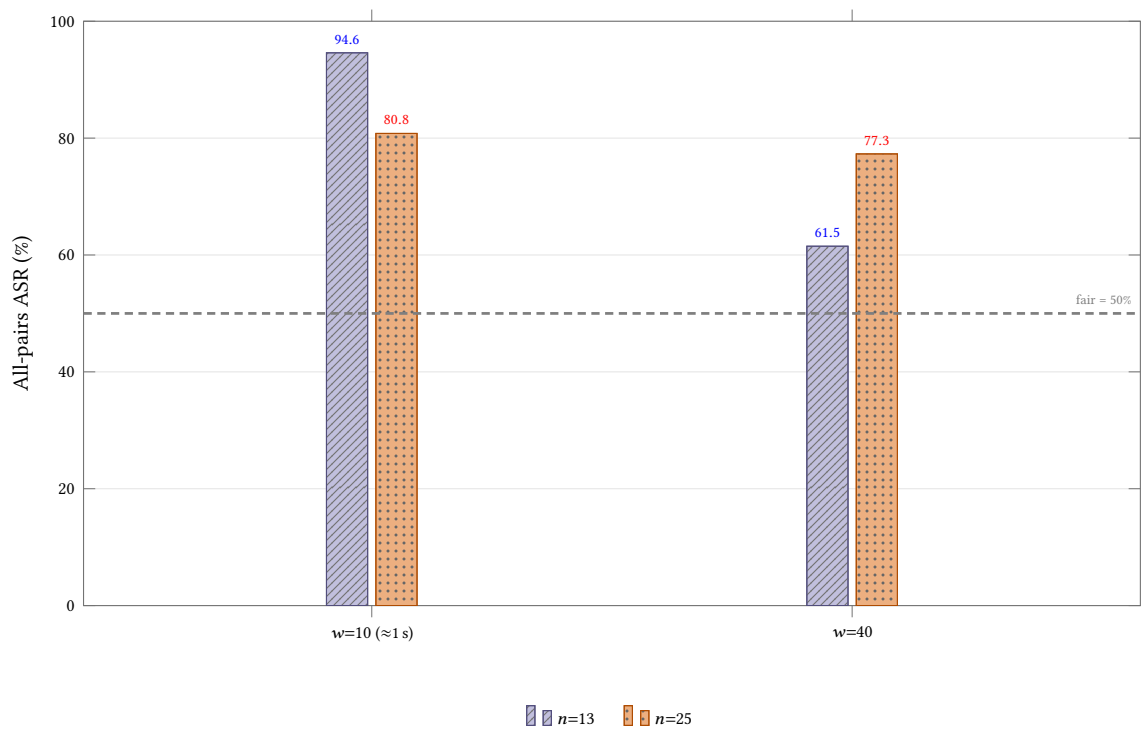
\begin{figure}[t]
  \centering
  \resizebox{\columnwidth}{!}{
\begin{tikzpicture}
\begin{axis}[
  suiBar,
  ybar=3pt, bar width=15pt,
  ylabel={All-pairs ASR (\%)},
  symbolic x coords={w10,w40},
  xtick=data,
  xticklabels={{$w{=}10$ ($\approx$1\,s)},{$w{=}40$}},
  x tick label style={font=\scriptsize, align=center},
  enlarge x limits=0.6,
  legend style={at={(0.5,-0.16)}, anchor=north, legend columns=-1, /tikz/every even column/.append style={column sep=0.3cm}},
]
\addplot+[draw=SuiPurple!70!black, fill=SuiPurple!45,
  postaction={pattern=north east lines, pattern color=black!55},
  nodes near coords, every node near coord/.append style={font=\tiny}]
  coordinates {(w10,94.6)(w40,61.5)};
\addplot+[draw=AccentOrange!80!black, fill=AccentOrange!50,
  postaction={pattern=dots, pattern color=black!60},
  nodes near coords, every node near coord/.append style={font=\tiny}]
  coordinates {(w10,80.8)(w40,77.3)};
\fairline{(w10,50)(w40,50)}
\node[font=\tiny, text=BaseGray, anchor=south east] at (rel axis cs:1,0.5) {fair = 50\%};
\legend{$n{=}13$, $n{=}25$}
\end{axis}
\end{tikzpicture}}
  \caption{Strategic silence keeps all-pairs ASR far above the 50\% fair line at both committee sizes. The short ($\approx$1\,s) window is the MEV-relevant
  one. Attacker stake fraction is held at $\approx$31\% across sizes.}
  \label{fig:silence}
\end{figure}

Figure~\ref{fig:silence} reports the result of strategic silence. At $n{=}13$, strategic silence lifts all-pairs ASR to $94.6\%$ over a 10-commit
window (roughly one second of mainnet activity, the timescale on which front-running and sandwiching actually happen) and $61.5\%$ over a 40-commit
window. Doubling the committee to $n{=}25$ (holding the attacker's stake fraction fixed) does not remove the effect: $80.8\%$ and $77.3\%$, respectively.

The gap narrows with larger committee size. This is expected because all-pairs ASR is dominated by cross-round pairs (at least $89\%$ of the pairs in these runs), which the attacker wins when its block lands in an earlier round than the victim's; because strategic silence front-loads the attacker's blocks, the score tracks how early that block-mass sits within the measurement window. At $n{=}13$ the attacker's blocks fill nearly the whole $10$-commit window (mean attacker round ${\approx}1.5$) but form only a thin early slice of the $40$-commit window (${\approx}16$ of $40$ rounds), so the advantage dilutes toward the $50\%$ fair line ($94.6{\to}61.5$). At $n{=}25$ the blocks instead stay in the front quarter of the window at both lengths (${\approx}3$ of $12$ vs.\ ${\approx}10$ of $41$ rounds), so the lead keeps pace with the window and the two scores nearly coincide ($80.8{\to}77.3$); the larger committee also lowers the short-window peak, because more validators dilute the attacker's dominance within any single round.

Note that these high rates assume an attacker controlling ${\approx}31\%$ of the stake, just under the Byzantine safety bound; a single validator with a realistic stake gains far less. Moreover, all-pairs ASR is a \emph{ratio} of pairs won, and strategic silence inflates it by producing \emph{fewer} but better-placed blocks. A high ratio, therefore, need not translate into more \emph{realized} profit, which depends on how many front-running slots the attacker actually fills, something our positional metric does not capture.
We also note that the same early-only block pattern can be reached through a bug rather than a choice: a validator that signs a far-future timestamp trips Sui's minimum-round-delay throttle (which uses a saturating subtraction) and is forced to skip rounds. That timestamp path relies on a missing validation (\S\ref{sec:fix}); strategic silence needs no such bug, which is why we treat it as the primary result.

\section{Is the Core Broken? A Fix}
\label{sec:fix}

It would be wrong to conclude that Mysticeti's ordering is generally unsafe. Below the BFT threshold, the standard in-protocol attacks (e.g., excluding a victim's blocks, delaying one's own proposals, withholding blocks from specific peers, or bribing a minority of validators) leave all-pairs ASR within about one point of 50\% in our runs. The commit rule still commits every honestly supported leader, and honest validators still advance rounds using only non-attacker blocks. The problems we report are not a failure of agreement; they are the consequence of one arbitrary \emph{tiebreak} (sorting equal-round blocks, and equal-fee transactions, by validator index) applied at the two points where the order is finalized.

This problem can be fixed by replacing the validator-index tiebreak with an \emph{unpredictable but deterministic} key. First, when linearizing a committed sub-DAG, sort equal-round blocks by a hash of the committed-leader digest and the block digest, instead of by author. Second, in the gas-price re-sort, break equal-fee ties by a hash of the leader digest and the transaction digest, rather than leaving them in consensus order. Both keys are fixed at commit time and identical for every validator, so the order stays deterministic (all nodes agree) and safety and liveness are untouched; both preserve the ``higher fee first'' rule that honest users rely on. Because the key depends on the leader digest, which a validator cannot grind, no index position is favored. Each change is confined to a single function. Encouragingly, Sui's own (currently gated) reputation-based leader schedule already replaced a fixed tiebreak with a per-epoch seeded shuffle for the same reason, so this style of fix is already considered acceptable in the codebase.

\begin{table}[t]
  \centering
  \footnotesize
  \setlength{\tabcolsep}{4pt}
  \begin{tabular}{@{}p{2.05cm}cp{2.0cm}p{1.9cm}@{}}
    \toprule
    \textbf{Flaw} & \textbf{Live?} & \textbf{Effect} & \textbf{Fix} \\
    \midrule
    Low-Index Head Start & yes & $\approx$89\% same-round win for low index & seeded per-commit tiebreak \\
    Tie Loophole & yes & bias reaches execution at tied fee, no premium & seeded gas-tie key \\
    Strategic Silence & yes (legit.\ action) & $>$94\% all-pairs over $\approx$1\,s & \emph{not} closed by tiebreak; needs round-level fix \\
    Unchecked Clock & yes (latent) & enables the timestamp variant & bound timestamp drift \\
    \bottomrule
  \end{tabular}
  
  \caption{Summary. Three exploits are active on Sui mainnet with no special configuration; one is present but currently low-impact. A single tiebreak
  change closes the head start and the tie loophole ($89\%\!\to\!45\%$ same-round); however, strategic silence needs a round-level mitigation.}
  \label{tab:flaws}
\end{table}

We implemented this tiebreak (env-gated; the default path is unchanged) and ran it on the $n{=}13$ cluster.
The head start collapses: same-round ASR falls from 89.3\% (median) to 44.8\%, i.e.\ to the 50\% fair line, confirming the validator index was its sole cause. Strategic silence, however, barely moves: its all-pairs ASR goes from 95.7\% to 92.3\%. The tiebreak removes silence's same-round component (100\% to 78\%) but not its dominant advantage, because silence wins by concentrating blocks in \emph{early rounds}, and the linearization is round-first out of causality: a block cannot precede its ancestors. Closing strategic silence, therefore, needs a round-level change, not a tiebreak, which we leave to future work.

The timestamp path behind the buggy variant of strategic silence exists because block verification never bounds a block's timestamp against the local clock. Today, this \emph{unchecked clock} has limited impact (the committed timestamp is a stake-weighted median that a minority cannot move), but
any future feature that reads a raw block timestamp would inherit an unchecked attack surface. A single drift-bound check at verification closes it.

\section{Conclusion}
\label{sec:conc}

Fair ordering on Mysticeti rests on two beliefs: that a DAG spreads ordering power across validators, and that a gas-price re-sort neutralizes what is left. We showed both are weaker than assumed and live on the Sui mainnet. First, linearizing the committed graph by validator index hands lower-indexed validators a permanent head start. Second, the gas-price re-sort, being a stable sort, does nothing on the tied fees that dominate real traffic, so the head start reaches execution with no premium paid; and third, a validator can amplify its position with the fully legitimate act of staying silent. None of this breaks consensus: below the Byzantine threshold, Mysticeti's across-round ordering remains fair, which is why a single change (replacing the index-based tiebreak with an unpredictable per-commit key) removes the structural head start (we measure the same-round bias fall from 89\% to the fair line) without touching safety, liveness, or fee priority. Strategic silence, which exploits round position rather than the tiebreak, is causality-bounded and remains the harder open problem. The broader lesson is that in a DAG protocol, the linearization tiebreak is not a harmless implementation detail: it is where fairness is actually decided.


\bibliographystyle{ACM-Reference-Format}
\bibliography{_blockchain,_system,_privacy,local_refs}

\end{document}